\documentclass[12pt]{iopart}
\input epsf
\begin{document}

\title[Bounds on the neutrino flux ... by Karl Mannheim]{
Bounds on the neutrino flux from cosmic sources of relativistic
particles
}

\author{Karl Mannheim
\footnote[3]{
Correspondence should be addressed to K.\,Mannheim
(kmannhe@uni-sw.gwdg.de)}
}

\address{Universit\"ats-Sternwarte, Geismarlandstrasse 11, D-37083 G\"ottingen,
Germany}

\begin{abstract}
In order to facilitate the identification
of possible new physics signatures in neutrino telescopes,
such as neutrinos from the annihilation of neutralinos
or decaying relics,
it is essential to gain
full control over the 
astrophysical inventory of neutrino sources in the Universe.
The total available accretion power,
the extragalactic gamma ray background, and the cosmic 
ray proton intensity can be used to constrain astrophysical models of 
neutrino production in extragalactic sources.  The resulting 
upper limit on the extragalactic muon neutrino
intensity from cosmic particle accelerators
$F_{\nu_\mu,\rm u.l.}\approx 10^{-6}$~GeV~cm$^{-2}$~s$^{-1}$~sr$^{-1}$ combined
with a reasonable minimum intensity of neutrinos due
to cosmic rays stored in clusters of galaxies 
$F_{\nu_\mu,\rm min}\approx
10^{-9}$~GeV~cm$^{-2}$~s$^{-1}$~sr$^{-1}$ 
demark a zone of opportunity for neutrino astronomy 
over a broad
range of energies between 100~MeV and 1~EeV.  Discovery of this neutrino
background 
would open a new era for astronomy and 
provide the first un-obscured view to the
early Universe.  

\end{abstract}

%\maketitle

\section{Introduction}
Surveying the sky with new instruments and methods
has many times prompted discoveries in the history of physics.
A well-known example is the sky survey obtained by Tycho Brahe
during 1576-1596,
unprecedented in terms of accuracy, scope, and sensitivity
at his time, which lead to Kepler's 1601-1619 discovery of the elliptical
planetary orbits and the three laws describing them.  This discovery
paved the way to Newton's 1664 theory of gravitation based on the notion of
an absolute metric space.
Another example is the work of Charles Messier who compiled
a catalogue of
nebular objects during 1771-1784 
to increase the finding probability for comets.
Edwin Hubble proved in 1925 
that among these sources were indeed 
galaxies gravitationally unbound to the
Milky Way initiating
extragalactic astronomy and 
observational cosmology.  \\

Electromagnetic waves, covering 
the electromagnetic spectrum from radio waves to gamma rays,
play the leading role in modern astronomy due to their
easy detection and diagnostic potential.
Other carriers of information are stable particles emitted from
cosmic sources, such as protons, ions, electrons, and neutrinos.
The cosmic ray protons, ions, and electrons
can be detected with great sensitivity, 
and studies of their
interactions in the Earth's atmosphere have
lead to a large number of discoveries in elementary particle physics,
the latest of which is the discovery of neutrino mass from the deficit
of large zenith-angle muon neutrinos (Fukuda et al.~1998). 
Cosmic rays are useless as astrophysical probes, however,
due to the omnipresence of interstellar and intergalactic fields
which 
deflect them away from the direction of the sources.
The exploration of the high-energy Universe
therefore relies on
radio astronomy, which is sensitive to the synchrotron radiation from
accelerated electrons, and gamma-ray astronomy, which probes
hadronic interactions of accelerated baryons and inverse-Compton
scattering and bremsstrahlung from accelerated electrons.  
However, the mean-free-path for gamma rays decreases to less
than the distance to the Galactic Center at energies above 
a few hundred TeV due to pair creation, 
and most of the Universe remains unseen.
Ultimately, photons do not penetrate the cosmic
photon-matter barrier making the first 10,000 years in the history of
the Universe invisible.\\

Neutrino astronomy (Gaisser et al.~1995,
Halzen 2000, Learned and Mannheim 2000)
can provide an un-obscured view to the Universe,
and neutrino detection is eased at
high energies due to

\begin{enumerate}
\item the increasing neutrino-matter interaction
cross section and muon range, 

\item the large natural water (or ice)
reservoirs transparent to the Cherenkov light produced in charged current
interactions, which can be used for neutrino-induced muon and shower detection, 

\item the steeply decreasing local foreground fluxes,
and 

\item the fascinating physics potential at high energies
(e.g., the neutralino with a mass of the order of 
the electroweak symmetry breaking
scale ${1 \over 2}(\sqrt{2}G_F)^{-1/2}\simeq 123$~GeV is one 
of the most favored candidates for the
prime component of cold dark matter and annihilations such as
$\chi \bar \chi \rightarrow W^+W^-,\, b\bar b$, and $Z\gamma$ would
lead to observable neutrinos).  
\end{enumerate}

However, until the time of writing, no 
extraterrestrial $>100$~MeV neutrino has been discovered presumably
due to the effective areas in the running experiments being still too small
for a detection.
In order to set the scale for experimentally reaching a zone of
discovery, 
theory attempts to provide a first estimate of the expected total
high-energy
neutrino intensity due to conventional astrophysical sources,
as has been done e.g.~by Stecker et al.~(1996), Mannheim (1995),
Protheroe \& Johnson (1996),
Waxman and Bahcall (1998), and Mannheim et al.~(2000).
Rarely in
the history of science has such a first estimate been close to reality.
For example, X-ray astronomy was considered a useless enterprise in the
50s, when it was known that stars had surface temperatures not exceeding
$10^4$~K.  With X-ray astronomy came the 
discovery of accreting stars, unexpected by most astrophysicists, 
which release gravitational rather than nuclear energy at an enormous
rate. Nevertheless, the attempt seems well justified, since a measured
flux greatly deviating from the predicted range of fluxes would
readily indicate a theoretical deficit prompting new developments.
In the following sections, possible routes to bounding the
allowed neutrino flux for models
of extragalactic neutrino production are outlined.

\section{Accretion power bound}

The dominant sources of relativistic particles in the Universe
are supermassive black holes in the centers of galaxies.
Since bright galaxies generally contain supermassive
black holes in their center with a mass proportional to the mass
concentrated in their stellar bulges (Rees and Silk 1998,
Gebhardt et al.~2000), it is possible to make a cosmic census
of all such black holes by using the well-known galaxy luminosity
function.  Moreover, since the mass in black holes was acquired
through accretion, one can
estimate the total
accretion power released in the Universe (Fabian 1999, Mannheim 1999).\\

Heavy elements with present-day mass fraction
$Z=0.03$ were produced in early bursts of star formation
by nucleosynthesis with radiative efficiency
$\epsilon=0.007$ yielding the present-day omnidirectional
stellar-light intensity
\begin{equation}
I_{\rm ns}\sim {c\over 4\pi}{\rho_* Z \epsilon c^2\over 1+z_{\rm f}}
\end{equation}
where $\rho_*$ denotes the mass density of baryonic matter
and $z_{\rm f}$ the formation redshift corresponding to
the era of maximum star formation 
(the ratio between
the energy released by stars and by other sources with the
same formation history is independent
of its details).
Let $\Omega_*$ denote
the baryon density in terms of the critical density
of the Universe and
$h=H_\circ/100~$km~s$^{-1}$~Mpc$^{-1}$ the
dimensionless Hubble constant, then
the intensity obtains the value
\begin{equation}
I_{\rm ns}\sim
1.4\times 10^{-2}\left(\Omega_*h^2\over 0.01\right)\left(1+z_{\rm f}\over
4\right)^{-1}\ {\rm GeV~cm^{-2}~s^{-1}~sr^{-1}},
\end{equation}
and this is in agreement with the observed intensity of
the integrated extragalactic background spectrum
from the far-infrared to the ultraviolet.  
Bright galaxies containing
supermassive black holes in their centers which are actively
accreting over a fraction of $t_{\rm agn}/ t_*\sim 10^{-2}$
of their lifetime 
produce an accretion-light intensity
\begin{equation}
I_{\rm accr}\sim {\epsilon_{\rm accr}M_{\rm bh}
\over  Z\epsilon M_*}
{t_{\rm agn}\over  t_*}I_{\rm ns}\sim 3.3\times 10^{-4}\ \rm 
GeV~cm^{-2}~s^{-1}~sr^{-1}
\end{equation}
adopting the accretion efficiency $\epsilon_{\rm accr}=0.1$
and the black hole mass fraction $M_{\rm bh}/M_*=0.005$.
Most of the accretion power emerges in
the ultraviolet where the diffuse background is unobservable
owing to photoelectric absorption by the neutral component of the
interstellar medium.  However, a fraction of $I_{\rm x}/I_{\rm bh}\sim
20\%$ taken from the average quasar spectral energy distribution
(Sanders 1989) shows up in hard X-rays
consistent with the observed hard X-ray background
bump 
$I_{\rm x}\sim 6.0\times 10^{-5}~\rm GeV~cm^{-2}~s^{-1}~sr^{-1}$
(Gruber 1992).
%\begin{figure*}
%\vspace{8cm}
%\centerline{\psfig{figure=figure1.ps,width=10cm,height=8cm}}
%\caption{Sketch of the present-day energy density
%of the extragalactic radiation background from radio waves to
%gamma rays. }
%\label{fig1}
%\end{figure*}
Non-thermal emission shows up
only in the radio-loud fraction $\xi_{\rm rl}\sim 20\%$ of
all AGN.  Among the radio-quiet AGN, hard X-ray emission is
common, but turns over steeply below 100~keV with no
signs for a nonthermal component.
The kinetic power of the jets in radio-loud AGN responsible
for the nonthermal emission roughly equals
the accretion power (Rawlings and Saunders 1991).  Hence
one obtains for the nonthermal intensity due to extragalactic jets
\begin{equation}
I_{\rm j}=\left(\xi_{\rm rl}\over 0.2\right)I_{\rm accr}\sim
6.7\times 10^{-5}\left(\xi_{\rm rl}\over 0.2\right)
\ \rm GeV~cm^{-2}~s^{-1}~sr^{-1}.                       
\end{equation}
This energy is released in relativistic particles, magnetic fields,
and pdV thermodynamic work against the ambient medium into which
the jets propagate.  
The gamma ray (cosmic ray, neutrino)
energy released by the jets amounts to
the present-day intensity 
\begin{equation}
I_\gamma\sim \xi_{\rm rad}I_{\rm j}\sim 6.7\times 10^{-6}
\left(\xi_{\rm rad}\over 0.1\right)\left(\xi_{\rm rl}\over 0.2\right)
\ \rm GeV~cm^{-2}~s^{-1}~sr^{-1}
\end{equation}
which is remarkably close to
the intensity $7.6\times 10^{-6}$~GeV~cm$^{-2}$~s$^{-1}$~sr$^{-1}$
of the extragalactic gamma ray background observed
between 100~MeV and 30~GeV
using the spectrum from Sreekumar et al.~(1998) and a
radiative efficiency of $\xi_{\rm rad}=10\%$.
Note that the intensity in the observed
gamma ray background is close
to the bolometric gamma ray intensity,  since pair attenuation and cascading
must lead to a turnover of the background spectrum above $20-50$~GeV
for extragalactic source populations
(Salamon and Stecker 1998). \\

It is therefore concluded, that 
if neutrinos tap a significant fraction of the
the total nonthermal accretion power available in the
Universe, then their bolometric intensity cannot exceed the intensity 
of the extragalactic gamma ray background, even
if gamma rays could not escape from the sources.\\

Since the observed black holes account precisely for the
extragalactic X-ray background, 
hidden supermassive black holes and neutrinos
from them (Berezinsky and Dokuchaev 2000)
are unlikely to exist:
they would have to lie
outside of bright galaxies, and they would have to be
heavily obscured at all times to avoid shining up in X-rays.
It is also possible to invoke a scenario
with hidden black holes which reside
in an advection-dominated accretion mode from which only
neutrinos escape.

\section{Extragalactic gamma ray background bound}

A more restrictive bound might apply for sources, from which
the gamma rays co-produced with neutrinos during pion decay escape.
The physical conditions inside typical 
nonthermal sources are characterized by a low matter density, indeed,
since this is the prerequisite for efficient
particle acceleration.
The observed gamma rays are likely to be the result of
electromagnetic cascading, taking place
during intergalactic travel or still inside the sources.
For typical extragalactic distances of the order of Gigaparsecs,
gamma rays above 100~GeV are expected to be absorbed
in collisions with low-energy photons from the metagalactic
radiation field produced by galaxies.  The resulting electron-positron
pairs re-radiate gamma rays by inverse-Compton scattering off microwave
background photons primarily in the MeV-to-GeV band, adding to
the primary radiation in this band.  
Therefore the observed extragalactic
gamma ray background intensity 
(Sreekumar et al.~1998)
represents an upper limit
to the maximum electromagnetic energy release due to pion
production
applying proper corrections for the kinematic branching between
neutrinos and gamma rays (Rachen and M\'esz\'aros 1998), 
e.g. $L_\gamma=2L_\nu$ for jets in active galactic nuclei
and $L_\gamma=L_\nu$ for jets in gamma ray bursts
(due to the different slopes of the target photon spectra).
Hidden sources, for which the escaping electromagnetic flux is
reprocessed thermally to below the 100~keV range, are possible,
but their total intensity must still remain below the accretion-related
bound outlined in the previous section - unless these hidden sources
would be related to an energy reservoir different from the supermassive
black holes in the centers of galaxies.

\section{Cosmic ray bound}

The bound discussed in the previous section applies to
extragalactic particle accelerators, from which
there is no contribution to the observed cosmic rays
due to escaping nucleons.
If, however, ultrarelativistic nucleons
escape and contribute to cosmic rays,
the observed cosmic ray intensity can be used to constrain models
of of neutrino production.\\

Even in sources, in which the accelerated particles are fully confined
(by magnetic fields), there is a flux of escaping neutrons produced
in the common isospin flip interactions.  Relativistic neutrons
at ultrahigh energies have decay lengths 
$\l_{\rm n}\simeq 10(\gamma_{\rm n}/10^9)$~kpc and leave 
the acceleration site and the host
galaxy without adiabatic losses.
The neutrino yield per escaping nucleon is highest
comparing with other escape mechanisms, e.g. 
it could be lower if there were additional prompt protons escaping
from the accelerator which are not accompanied by neutrinos.
By the logic inherent to an upper limit, the discussion is restricted
here to the case of neutron-origin cosmic rays.  \\

An extragalactic flux
of cosmic rays from evolving sources would be mostly protonic due to the
photo-disintegration of heavy nuclei (Stecker and Salamon 1999).
The observed
upper limit on extralactic protons can be converted to an upper limit
on extragalactic neutrinos by virtue of the production and decay
kinematics.  This has been worked out in detail by Mannheim, Protheroe,
and Rachen (2000) using a propagation code developed by Protheroe and
Johnson (1996) and assuming an evolution of the comoving emissivity
of the putative cosmic ray/neutrino sources given by
\begin{equation}
(dP_{\rm gal} / dV_c)\langle Q(E,z)\rangle\propto (1+z)^{3.5},
\end{equation}
which has the same redshift dependence
as the cosmic star formation rate 
or the AGN emissivity (Boyle and Terlevich 1998).
The cosmic ray intensity at Earth at
energy $E$ due to neutron decay 
is given by
\begin{equation}
\label{CRint}
I(E)  \propto
 \int\limits_{z_{\rm min}}^{z_{\rm max}} \!\!\! M(E,z)
 {(1+z)^2 \over 4 \pi d_L^2} {{\rm d} V_c \over {\rm d}z}
 \frac{{\rm d}P_{\rm gal}}{{\rm d}V_c}
 \big\langle Q[(1+z)E,z]\big\rangle \, {\rm d}z
\end{equation}
where $d_L$ and $V_c$ are luminosity distance and co-moving volume, and
$M(E,z)$ are ``modification factors'' for injection of protons at redshift
$z$ as defined by Rachen and Biermann (1993); for neutrinos,
$M(E,z)=1$.  The modification factors for protons depend on the input spectra,
and are calculated numerically using a matrix method (Protheroe
and Johnson 1996).   
The trial functions for the spectra emitted by extragalactic 
photo-production sources with target spectral index $\alpha=1$
(typical for conditions in jets from active galactic nuclei)
are
\begin{equation}
Q_{\nu_\mu}(E)=83.3 Q_{\rm n}(25 E)
\end{equation}
and 
\begin{equation}
Q_{\rm n}(E)\propto E^{-1} \exp[-E/E_{\rm max}]
\end{equation}
with variable $E_{\rm max}$.  The construction of the upper limit
now proceeds by first choosing a value for $E_{\rm max}$ and
finding the maximum source emissivity which does not overproduce 
either the
experimental upper limit on cosmic ray protons 
or the observed extragalactic gamma ray
background (comparing integrated fluxes because of electromagnetic
cascading which destroys the proportionality between the differential
spectra).  Second, the neutrino intensity is computed for the
maximum allowed emissivity at $E_{\rm max}$, and third, the
envelope curve for a set of $E_{\rm max}$ values then defines
the upper limit (see Mannheim et al.~2000 for more details).
The resulting upper limit is largely robust with respect
to the specific spectral shape generated by a source distribution.\\

Below neutrino energies of $\sim 10^5$~GeV the constraint due to the
observed cosmic rays becomes sharper than that due to the extragalactic gamma
ray background. The upper limit decreases with energy
reaching a minimum at around $\sim 10^9$~GeV.  At higher energies,
the upper limit increases for technical reasons:
the observed cosmic rays show an
excess above the so-called GZK cutoff which would require an emissivity
increasing with energy in order to compensate the photo-hadronic
energy losses due to propagation through the microwave background.

\section{Neutrinos due to cosmic ray storage in clusters of galaxies}

It was shown in Sect.~2 that the intensity in relativistic
particles available through accretion is 
$\sim 6.7\times 10^{-5}$~GeV~cm$^{-2}$~s$^{-1}$~sr$^{-1}$.
The supermassive black holes giving rise to radio jets preferentially
lie in clusters of galaxies which contain an intracluster medium
of gas density $n_{\rm c}\approx 10^{-3}$~cm$^{-3}$ which acts as
a target for the relativistic particles from the 
jets diffusing through
the intracluster medium. The optical depth of this medium with
respect to pion production is $\sim 9\times 10^{-5}$, and, using
$I_\gamma=I_\nu$, 
the corresponding minimum neutrino intensity due to cosmic ray
storage is of the order of 
$I_\nu\approx 3\times 10^{-9}$~GeV~cm$^{-1}$~s$^{-1}$~sr$^{-1}$.
Outside of their sources, the stationary proton distribution should
reflect the source distribution.  For relativistic shocks, this distribution
is expected to be $dN/dE\propto E^{-2.2}$ (Kirk et al.~2000), 
and the lower limit in
Fig.~1 is normalized by the bolometric condition
$\int_{1}^{10^{11}} E dN=I_\nu$ ($E$ in GeV).
The corresponding
value $I_\gamma=I_\nu$ is consistent with upper limits on the
gamma ray flux from clusters of galaxies and is in agreement with the
much more detailed model of Colafrancesco and Blasi (1998).

\begin{figure}
\begin{center}
{\centerline{\epsfxsize=12cm
\epsfbox{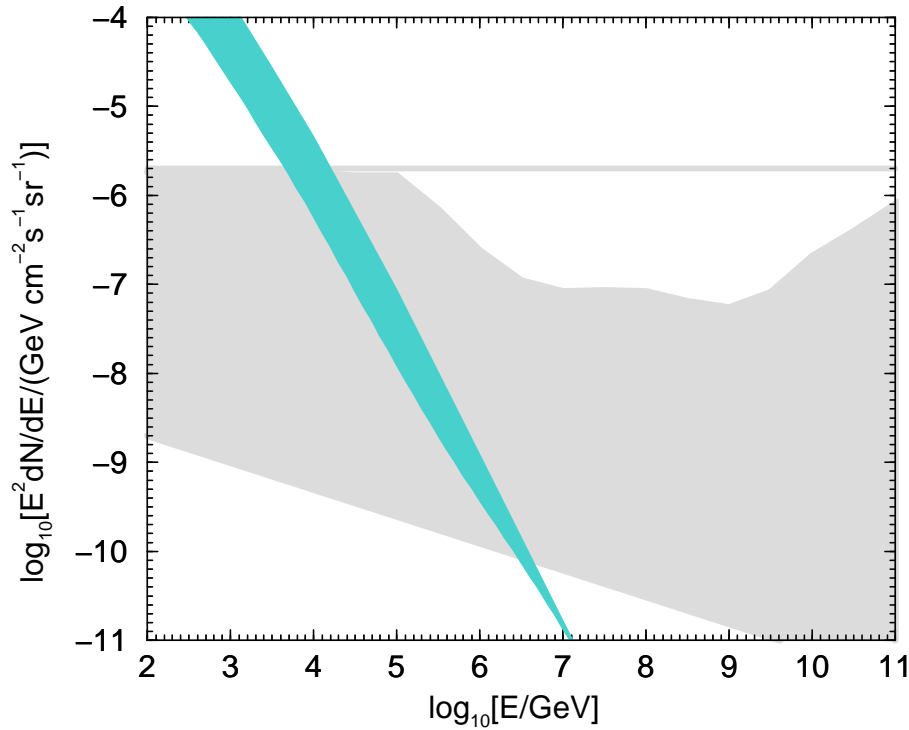}}}
\end{center}
\caption{Upper limits on the extragalactic neutrino intensity based on
(i) the extragalactic gamma ray background
({\it straight grey line})
and (ii) the observed cosmic ray intensity ({\it
upper edge of gray shaded area}) ignoring the possible effects of large-scale
magnetic fields. The {\it lower edge of the gray
shaded area} depicts the lower limit due to cosmic ray storage
in clusters of galaxies assuming that the nonthermal accretion power
is channeled into a power-law
distribution of protons. For comparison, the figure also shows
the intensity of atmospheric neutrinos for directions from
vertical to horizontal,
including the omnidirectional contribution from prompt charm production
estimated by Thunman et al.~(1996)
({\it dark shaded area}).}
\end{figure}

\section{Discussion}

The presence of  high-energy cosmic rays in extragalactic sources is
inevitably  associated with gamma rays and neutrinos.  In principle, both
cosmic ray nucleons and gamma rays could be absorbed in hidden sources, 
observable only in neutrinos. However,   the observed gamma-ray background 
can be attributed to the  nonthermal fraction of  the gravitational binding
energy released during accretion of ambient matter onto supermassive black
holes, and there is no other known energy reservoir available for hidden
sources.  If these putative sources tap a fraction of the nonthermal accretion
energy of supermassive black holes, then this fraction must be very small, or
there must exist heavily
obscured supermassive black holes in great numbers outside of galactic
centers which have escaped the cosmic census so far.
Thus, an intensity in excess
of the accretion bound would indicate this new population of black holes, or
might indicate a new weak-channel dissipation process of the cold dark matter
in the Universe.\\

The observed extragalactic gamma ray background is probably due to
unresolved faint gamma ray point sources (associated with rotating
supermassive black holes), since the known
$\sim 100$ resolved sources above 100~MeV alone
produce a cumulative flux 
amounting to $\sim 0.15$ of the total diffuse flux.
We do not know much about the opacity for gamma
rays in the sources, but their spectra generally do not seem
to indicate a spectral turnover up to GeV energies.
This translates into an opacity constraint for protons with
respect to the photo-production on the same target photons
and implies that these
sources should typically be thin to the emission of 
cosmic ray nucleons up to 
$\sim 10^8$~GeV (corresponding to neutrino energies $\sim 10^6$~GeV).
Since the optically thin limit approaches the gamma ray limit
at energies of $\sim 10^5$~GeV, just a little effect due to large-scale
magnetic fields raising the transition energy
between the two limits to $\sim 10^6$~GeV (as argued in Mannheim et
al.~2000, chapter V.) would likely render the gamma ray limit as the
only conservative upper limit over the entire energy range.
Future gamma ray observations could enhance the
relevance of the cosmic ray bound, if it can be shown
that the gamma ray spectra show no signs of intrinsic absorption
up to energies much higher than GeV.\\

Some sources have been detected up to TeV  without showing
signs of internal absorption (albeit the steep multi-TeV spectra could
indicate pair creation above TeV).  These sources could be strong sources of
neutrons (protons) up to $10^{11}$~GeV and the observed cosmic ray flux provides
an upper limit to the allowed neutrino flux at $\sim 10^{9}$~GeV 
almost two orders
of magnitude below the gamma ray limit.  As shown in Mannheim (1995),
a source population such as BL Lacertae objects, likely to produce
a distribution of $E_{\rm max}$ values, and saturating the cosmic
ray bound at ultrahigh energies would nevertheless produce a
significant fraction of the entire extragalactic gamma ray
background  (the bolometric gamma ray bound in
a differential plot may be misleading in this case,
since it is contructed for very hard trial spectra).
\\

No gamma ray source is known to be definitely thin to the emission of cosmic
rays above $10^{11}$~GeV.  In principle, however, such a population could exist
and could
provide hard spectra of cosmic
rays compensating the photo-production energy losses during passage of the
microwave background with an emissivity strongly increasing with redshift.  In
this very unlikely
case, the rising cosmic ray bound technically
reflects the fact, that an excess of
cosmic ray events has been found above the so-called GZK cutoff energy of
$5\times 10^{10}$~GeV.\\

\section{Conclusions}

There are two arguments in favor of an extragalactic high-energy neutrino
background bounded from above by the known intensity of the
extragalactic gamma ray background and from below by the known matter
column density
of clusters of galaxies:  naturalness and the likelihood of an
extragalactic origin
of the ultrahigh energy cosmic rays.  It is natural to assume that
a significant fraction of the nonthermal power produced
by black holes is channeled
into protons, since the transport of electrons
out of the compact source region is strongly damped by radiative losses.
The observed near isotropy and protonic character of the extended air
showers at ultrahigh energies argues in favor of an
extragalactic origin, and the gamma ray spectra from extragalactic jets
are in accord with the assumption that protons are accelerated
in them to ultrahigh energies
(albeit there are competing explanations based on the assumption
of electron acceleration, for thorough
discussions see Mannheim 1998 and Wilson et al.~2000).  
The picture certainly needs experimental clarification by
showing that extragalactic neutrinos with an
intensity marked by the ''zone of discovery'' 
(the grey zone in Fig.~1) really exists.  
If a stronger intensity were found, this
could imply a new population of 
completely obscured black holes outside of bright galaxies
or a signature of relics
from the Early Universe.  If a weaker intensity were found, then
extragalactic jets would not produce the ultrahigh energy cosmic rays,
and accelerated protons would not be responsible for the observed
gamma rays.  

\ack

I thank
Raymond Protheroe and J\"org Rachen for their invaluable contributions to
this research, and Wlodek Bednarek for stimulating discussions (in
particular for pointing out the possible relevance of advection-dominated
blackk hole accretion).
Support by the Deutsche Forschungsgemeinschaft is
gratefully acknowledged.

\section*{References}
\begin{harvard}

\item Berezinsky VS, Dokuchaev VI  2000 {\it Astropart. Phys.}, in press (astro-ph/0002274)
\item Boyle RJ, Terlevich RJ 1998 {\it Mont. Not. Roy. Astro. Soc.} {\bf
293} L49
\item Colafrancesco S, Blasi P 1998 {\it Astropart. Phys.} {\bf 9} 227
\item Fabian, AC 1999 {\it Mon. Not. Roy. Astr. Soc.} {\bf 308} 39
\item Fukuda Y, et al.~(The Super-Kamiokande Collaboration)
1998 {\it Phys. Rev. Lett.} {\bf 81} 1562
\item Gaisser TK, Halzen F, Stanev T, {\it Phys. Rep.} {\bf 258} 173
\item Gebhardt, K et al.~2000 {\it Ap.J.} {\bf 539} L13
\item Gruber DE 1992 in: {\it The X-ray Background}, eds. X Barcons and 
AC Fabian, Cambridge UP, p.~44
\item Halzen F 1999 {\it Astropart. Phys.} {\bf 2} 88
\item Halzen 2000 {\it Phys. Rep.} {\bf 333-334} 349
\item Kirk JG, Guthmann AW, Gallant YA, Achterberg A 2000 {\it Ap. J.} {\bf
542} 235
\item Learned J, Mannheim K 2000 {\it Ann. Rev. Nucl. Part. Sci.} {\bf 50}
679
\item Mannheim K 1995 {\it Astropart. Phys.} {\bf 3} 295
\item Mannhein K 1998 {\it Science} {\bf 279}, 684
\item Mannheim K 1999 {\it Astropart. Phys.} {\bf 11} 49
\item Mannheim K, Protheroe RJ, Rachen JP 2000 {Phys. Rev. D} accepted
(astro-ph/9812398)
\item Protheroe RJ, Johnson 1996 {\it Astropart. Phys.} {\bf 4} 253
\item Rachen JP, Biermann PL 1993 {\it Astron. Astrophys.} {\bf 272} 161
\item Rachen JP, M\'esz\'aros P 1998 {\it Phys. Rev. D} {\bf 58} 123005
\item Rawlings S, Saunders S 1991 {\it Nature} {\bf 349} 138
\item Rees MJ, Silk J 1998 {\it Astron. Astrophys.} {\bf 331} L1
\item Salamon MH, Stecker FW 1998 {\it Ap. J.} {\bf 493} 547
\item Sanders DR 1989 {\it Ap. J.} {\bf 347} 29
\item Sreekumar P, et al.~(The EGRET Collaboration) 1998 {\it
Ap.J.} {\bf 494} 523
\item Stecker FW, Salamon MH 1996 {\it Space Sci. Rev.} {\bf 75} 341
\item Stecker FW, Salamon MH 1999 {\it Ap. J.} {\bf 512} 521
\item Thunman M, Ingelman G, Gondolo P 1996 {\it Astropart. Phys.} {\bf 5} 309
\item Waxman E, Bahcall JB {\it Phys. Rev. D} {\bf 59} 023002
\item Wilson AS, Young AJ, Shopbell PL 2001 {\it Astrophys. J.} {\bf 546},
Jan 10 issue (astro-ph/0008467)

\end{harvard}

\end{document}